\documentclass[acmsmall]{acmart}
\AtBeginDocument{%
  \providecommand\BibTeX{{%
    Bib\TeX}}}
    
\setcopyright{acmlicensed}
\copyrightyear{2025}
\acmYear{2025}
\acmDOI{XXXXXXX.XXXXXXX}

\usepackage{amsfonts}
\usepackage{algorithmic}
\usepackage{textcomp}
\usepackage{tabularx}
\usepackage{enumitem}
\usepackage{colortbl}
\usepackage{natbib}
\usepackage{listings}
\usepackage{wrapfig}

\def\BibTeX{{\rm B\kern-.05em{\sc i\kern-.025em b}\kern-.08em
    T\kern-.1667em\lower.7ex\hbox{E}\kern-.125emX}}
\usepackage{subcaption}

\usepackage[most]{tcolorbox}

\begin{document}

\setcopyright{cc}
\setcctype{by-nc-nd}
\acmJournal{FACMP}
\acmYear{2026} \acmVolume{1} \acmNumber{1} \acmArticle{}

\title[Advancing SE through the Lens of Causation]
{Reasoning Beyond Prediction: From Data-Driven to Causal Software Engineering}

\author{Roberto Pietrantuono}
\email{roberto.pietrantuono@unina.it}
\orcid{0000-0003-2449-1724}
\author{Luca Giamattei}
\email{luca.giamattei@unina.it}
\author{Stefano Russo}
\email{stefano.russo@unina.it}
\affiliation{%
  \institution{University of Naples Federico II}
  \city{Naples}
  \country{Italy}
}

\renewcommand{\shortauthors}{Pietrantuono et al.}

\begin{abstract}

 Software engineering is an intellectually demanding, creative discipline that juggles a web of interdependent tasks to design, build, and assure the quality of increasingly complex systems. As our expectations from software soar — with demands spanning AI-driven products, pervasively distributed and cloud-native architectures,
 and deeply embedded cyber-physical environments — its complexity steadily increases.
 In response, a new wave of co-engineering methods and tools, fueled by deep learning, has emerged to augment the process, enhancing automation and decision support. Yet, these advances remain far from delivering the kind of intelligent support that modern software development demands. We call for a new paradigm of human–machine cooperation: one where machines don't just automate routine tasks or predict from learned patterns, but actively amplify engineers' \textit{reasoning} through the lens of \textit{causation}. 
 As software becomes smarter, a smarter support is needed.   %

\end{abstract}



\received{\today}

\maketitle
\section{Introduction}

Over the past 10–15 years, many software systems have shifted toward cloud-native and distributed architectures, although adoption remains heterogeneous across organizations and domains. Industry surveys report widespread adoption of cloud-native practices (e.g., 89\% in 2024) and broad production use of Kubernetes — both strong indicators of large-scale distributed deployment \cite{CNCF2024,CNCF2023}. In parallel, modern systems increasingly incorporate adaptive and data-driven components, including AI/ML in many contexts \cite{McKinsey2024}, enabling capabilities such as autonomous vehicles, intelligent robots, LLM-based assistants, and virtual doctors.
But these advances also bring increased uncertainty and unpredictability, driven by data-centric computation, intricate interactions with dynamic environments, and the pervasive distribution of today’s architectures. 
Microservice ecosystems may include hundreds of independently deployable services; cloud-native applications run atop dynamic infrastructures; cyber-physical systems must coordinate with uncertain physical environments; and AI/ML components can introduce probabilistic behavior and novel failure modes. Together, these factors contribute to structural and epistemic complexity beyond that of traditional monolithic or statically deployed systems — as evident, for instance, in microservice architectures \cite{TemporalMicroservicesTD} and AI/ML-enabled systems \cite{Martinez22}. This shift toward highly interconnected, adaptive, and data-driven architectures makes it increasingly difficult to reason about global behavior, safety, and correctness, requiring engineers to navigate vast behavioral spaces to uncover critical failures. 
The question is no longer \textit{Can we build it?} — but \textit{Can we trust it?} Failing to answer 
could seriously erode public confidence in future systems, with unpredictable social and economic impact.\footnote{For a deeper discussion of how trust is conceptualized in software engineering and AI assistants, see \cite{khati2025,baltes2025_tse}}

\textit{But how are software engineers managing this challenge?} %
 Under the banner of AI4SE  (Artificial Intelligence for Software Engineering), deep learning-powered human–machine co-engineering — now most visibly through LLM-assisted development \cite{Hou2024} — 
 is constantly advocated as \textit{the} solution.  
Undoubtedly, the gains are real and sometimes impressive: 
today's tools can autocomplete code, draft tests, generate documentation, and even offer design suggestions. 
 However, the steady stream of reported, sometimes catastrophic, software-related incidents
 shows that we are still far from 
 trustworthy software systems. %
 In practice, software engineering struggles to keep pace with demands for safety, fairness, security, privacy, and transparency, to name just a few concerns. The reason is structural, as the inherent flaws of deep learning — rooted in the correlation-based nature and lack of transparency — makes AI support fragile, opaque, and hence untrustworthy, especially in critical domains. Unsurprisingly, %
 the reliability and quality of LLMs themselves has emerged as one of the hottest research areas in 2024-25 \cite{tie2024_tosem, Battaglini2025}. In the end, ML- and LLM-powered solutions suffer from the same problems they are supposed to solve.

\textbf{A simple example illustrates the gap.}
Consider an autonomous driving system, like a city shuttle. During peak hours, the shuttle occasionally brakes too late at pedestrian crossings. Logs appear normal, and an ML-based monitoring tool flags a strong correlation: \textit{late braking (Y) tends to occur when the shuttle is using high-resolution camera mode (X)}. Based on correlation alone, the tool blames the camera pipeline. But this is a classic case of spurious correlation.  A causal analysis reveals the true driver: higher vehicle speed (Z) automatically triggers the switch to high-resolution camera mode (X) to improve long-distance perception, adding processing delay; the same increase (Z) also reduces the available braking margin, making late braking (Y) more likely. The system incorrectly blames X for Y, when both are caused by Z.  A causal model makes this explicit and also allows engineers to pose targeted \textit{what-if} questions based on these relations  (e.g., What if the vehicle enters a downhill segment? What if road friction changes? What if we adjust the speed threshold for sensor-mode switching?) to identify risky configurations before deployment.  Such reasoning reveals risks that today’s correlation-based tools simply cannot detect.  This small example captures a broader reality: correlation-based tools can tell us what happened, but rarely \textit{why}, nor \textit{what would happen if} things were different. And without the why, trust remains elusive.
We believe it is time for a \textit{qualitative leap} in the kind of assistance machines can offer to engineers: AI should \textit{not} simply do more; it should do it better.

\vspace{-3pt}
\section{Human-Machine Software co-Engineering}

\textit{Machines} have long supported \textit{humans} by automating tedious and error-prone parts of engineering processes, including \textit{software engineering} (SE). 
Over time, their assistance has evolved, accelerating both development and quality assurance. In software engineering, this progression can be seen in \textit{what} machines support (Sidebar 1): from ancillary, problem-independent tasks and scaffolding (e.g., CASE tools, workbenches, IDEs) (row 1), to automating techniques that formalize engineers’ intuitions in analysis, design, implementation, testing, and maintenance (row 2). For example, Coverage-Guided Testing formalizes the intuition that \textit{increasing test coverage reveals more failures}; Automated Static Analysis formalizes rule-checking to detect code quality issues. The rise of \textbf{machine learning} (ML) — particularly deep learning — has been indeed a game changer for SE (row 3). With ML, the machine is asked not only to \textit{automate tasks} but also to \textit{support decisions} by exploiting historical data from the system or process to uncover patterns across the lifecycle.

This paradigm -- known as \textbf{Data-driven Software Engineering} (DDSE) — has elevated the role of the machine in  human-machine co-engineering, as ML augments human capabilities and participates in decision-making. Today, ML is used to support nearly every SE activity \cite{Wang2023}. 
Companies deploy ML models to predict (mis)behaviors by learning correlations from product, process, and operational data across the lifecycle. In IT operations, this is often framed as AIOps, an ML-driven approach that automates operations by detecting patterns in data and triggering remediation.%

As \textbf{LLMs came to the scene}, a new wave of data-driven solutions is emerging. %
LLMs utilize vast unstructured datasets (e.g., from the web), acting as surrogates of domain expert knowledge to enhance tasks such as requirements analysis \cite{huang2025}, coding \cite{jiang2024},  testing \cite{Wang2024}, debugging \cite{Majdoub2024}, repair \cite{zhang2024_tosem}, %
and roughly any software engineering task \cite{Hou2024}. 
Despite the challenges and concerns surrounding LLM-based solutions (e.g., safety and privacy risks), companies are increasingly adopting  them,\footnote{\url{https://www.mckinsey.com/capabilities/quantumblack/our-insights/the-state-of-ai}}$^,$\footnote{\url{https://www.statista.com/outlook/tmo/artificial-intelligence/worldwide}} although the extent of the actual benefits remains a debated topic.\footnote{\url{ https://www.heise.de/en/news/Warnings-of-the-AI-bubble-bursting-are-increasing-10767973.html}}$^,$\footnote{\url{https://www.gartner.com/en/newsroom/press-releases/2025-06-25-gartner-predicts-over-40-percent-of-agentic-ai-projects-will-be-canceled-by-end-of-2027 }}
DDSE represents the current state of human–machine co-engineering. Compared to two decades ago, machines now do far more than automating repetitive tasks. Their step-beyond has been \textit{qualitative}: ML scales and automates a core human ability, \textit{pattern recognition}. By uncovering patterns in large datasets, ML amplifies \textit{inductive reasoning}, enabling systems to infer complex input–output relations and make predictions that guide decisions. LLMs extend this with content generation based on learned patterns. How accurate these capabilities are, and how much engineers should rely on them, is a separate debate, but it is undeniable that, when applied properly, ML can boost software engineering productivity by an order of magnitude.

\tcbset{
  sidebar style/.style={
    enhanced,
    colback=cyan!3,
    colframe=blue!3,
    width=1\textwidth,
    breakable,
    boxsep=2pt,         %
  left=1pt,           %
  right=2pt,          %
  top=10pt,            %
  bottom=4pt,         %
    boxrule=0.5pt,
    sharp corners,
    fontupper=\footnotesize,%
    title={Sidebar 1: Machine Support in Software Engineering: Key Milestones},
    attach boxed title to top center={yshift=-2mm},
    boxed title style={colback=blue!50, colframe=blue!40}
  }
}
\begin{tcolorbox}[float*=t,sidebar style]%
\renewcommand{\arraystretch}{1.5}
\resizebox{\textwidth}{!}{%
\begin{tabularx}{\textwidth}{p{2,3cm}p{4,4cm}X}
\cellcolor{blue!50}\textbf{\textcolor{white}{Goal}} & \cellcolor{blue!50}\textcolor{white}{\textbf{Description}} & \cellcolor{blue!50}\textcolor{white}{\textbf{Examples}} \\

Scaffolding & Tools provide general, problem-independent support for organizing and structuring engineering activities. & CASE tools, IDEs, workbenches; syntax highlighting; version control systems; build automation; test runners; basic linting tools. \\

Support automation & Machines implement formalized methods and techniques to automate specific software engineering tasks. & Coverage-Guided Testing (CGT); Automated Static Analysis (ASA); Model Checking; Symbolic Execution; Code Generation from Models. \\

Support decision & ML-based tools assist decision-making by learning from historical data, recognizing patterns, making predictions. & Effort estimation; code completion; fault prediction; test generation, selection \& prioritiz.; RCA; 
anomaly detection; failure prediction; doc generation. \\

\cellcolor{cyan!10} \textbf{Support reasoning} & \cellcolor{cyan!10}  Machines perform step-by-step reasoning to support explanation, hypothesis generation, and action planning. & \cellcolor{cyan!10} What-if analysis; interactive simulation and diagnosis; counterfactual reasoning; explainable recommendations; RCA. \\
\end{tabularx}
}
\label{sidebar1}
\vspace{-6pt}
\end{tcolorbox}
\vspace{-6pt}

\section{The step beyond: Causal Software Engineering}

Although ML, and especially LLMs, is often perceived as a panacea, it supports only part of the engineering process. While learning patterns in data is essential for prediction, engineers do much more when designing or validating a system. Throughout the development cycle, they formulate and compare alternative hypotheses, mentally simulate actions, and seek explanations — in other words, they \textbf{reason} beyond pattern recognition. 
For example, engineers \textit{hypothesize} multiple design options to meet requirements. During quality assurance, they anticipate how the system might behave under specific conditions and ``guess'' which inputs, configurations, environmental factors, or user behaviors could trigger failures. In maintenance, such as debugging, they generate alternative explanations for \textit{why} a failure occurred and what might have happened under different scenarios, guiding them toward an effective fix.
In each of these tasks, engineers naturally infer \textbf{causal relationships} among variables and, by such relations, mentally simulate alternative scenarios. We envision causal reasoning as a promising direction for enhancing next-generation assistants.

ML models — including \textit{transformers}, the foundation of LLMs — primarily learn \textit{associations} between variables, corresponding to  rung 1 of Pearl’s ladder of causation \cite{Pearl18}. They do not uncover \textit{causality}. With ML, we can only \textit{reason about what has been observed}, answering questions such as: given an observed context (from which the model learned), \textit{what output $X$ should we expect if the input $W$ happens to be $w$?} Formally, this is the conditional probability $P(X \mid W = w)$. Even with LLMs and their massive training corpora, the output fundamentally remains a conditional probability over observed patterns.
Causal reasoning acts at a higher level: it targets causation and answers questions such as \textit{What output $X$ do we expect if we \underline{actively set} $W = w$?} — that is, if we intervene on $W$ and change the data-generating process. Or: \textit{What would have happened to $X$ if we \underline{had set} $W = w$?} 
These \textbf{intervention} and \textbf{counterfactual} queries (rungs 2 and 3 of the ladder of causation) lie beyond ML's reach, since ML predicts only from what has been seen. Causal reasoning, by contrast, allows us to \textit{proactively} explore what will happen (or would have happened) under hypothetical interventions, rather than waiting for data to appear. It shifts from a \textit{learning-by-seeing} paradigm toward \textit{learning-by-doing} or even \textit{learning-by-imagining}. Thus, machines gain not only the \textit{learn-to-predict} capability of ML, but also a \textit{plan-to-improve} capability. 

 The growing importance of 
 reasoning is evident in the evolution of LLMs from advanced yet associative ML algorithms to compositional systems that demonstrate increasingly effective forms of \textit{approximate reasoning}, driven by techniques such as chain-of-thought prompting, self-refinement, and reasoning-oriented training. The push to integrate LLMs with external tools to support complex reasoning tasks — seen in AI agents and virtual assistants — points in the same direction. 

The scientific community is actively debating the extent to which LLMs genuinely reason, as opposed to merely exhibiting the appearance of reasoning through sophisticated pattern matching such as memorization effect \cite{Plaat25}. 
At the same time, a growing body of work examines the synergy between \textit{LLMs and causal reasoning}, where causal models provide the structure for formal reasoning and LLMs support causal learning and inference \cite{kiciman2024, wan2025, liu2025}. In domains such as safety engineering, law, and healthcare — where explanation, justification, and trust are essential — equipping LLMs with causal reasoning is especially important. Without such a shift, the impact of AI in software engineering may plateau as demands for explainability and accountability grow sharper. %

\textbf{Causal Software Engineering} (CSE) offers a new human–machine co-design perspective in which machines \textit{imitate} and \textit{augment} human reasoning to anticipate scenarios and explain events, amplified by computational power. Given the motivations above,
we believe CSE may represent the next qualitative step beyond DDSE in the assistance machines can provide. The rationale lies in how information is used: DDSE implicitly assumes that \textit{data is all you need}, yet — no matter how much data you have — useful answers and genuine reasoning require a model where \textit{human knowledge} and \textit{assumptions} are explicit. Statistical causal modeling is one way (though not the only one) to harness data to answer \textit{the right} questions and filtering out spurious correlations.\footnote{While various forms of automated logical reasoning are possible, we focus on statistical (covariance-based) causal reasoning \cite{kiciman2024} as a natural step beyond ML \cite{Pearl18}. This approach leverages both domain knowledge (encoded in a model) and the rich data available in SE processes.} Sidebar 2 highlights the main differences between DDSE and CSE.

\tcbset{
  sidebar style/.style={
    enhanced,
    colback=cyan!3,
    colframe=blue!3,
    width=1\textwidth,
    breakable,
    boxsep=2pt,         %
  left=1pt,           %
  right=2pt,          %
  top=6pt,            %
  bottom=4pt,         %
    boxrule=0.5pt,
    sharp corners,
    fontupper=\footnotesize,%
    title={Sidebar 2: Data-Driven Software Engineering vs Causal Software Engineering},
    attach boxed title to top center={yshift=-2mm},
    boxed title style={colback=blue!50, colframe=blue!40}
  }
}

\begin{tcolorbox}[float*=t,sidebar style]%
\renewcommand{\arraystretch}{1.5}
\resizebox{\textwidth}{!}{%
    \begin{tabularx}{\textwidth}{XX}

 \multicolumn{1}{c}{\textbf{\textcolor{blue!70}{Present -- DDSE based on ML}}} &  \multicolumn{1}{c}{\textbf{\textcolor{blue!70}{Future -- CSE based on Causal Reasoning}}}\\

\cellcolor{blue!50}\textbf{\textcolor{white}{Pattern Recognition}} & \cellcolor{blue!50}\textbf{\textcolor{white}{Causal Inference}}\\ 
ML automates and amplifies our pattern recognition ability, allowing us to uncover unknown associations from data. 
& 
Causal relations – implying associations - enable richer reasoning tasks such as simulating what-if scenarios, deriving explanations, formulating hypotheses, and imagining alternative pasts. Automating these abilities substantially elevates the quality of machine assistance.
\\
\cellcolor{blue!50}\textbf{\textcolor{white}{Correlation}} & \cellcolor{blue!50}\textbf{\textcolor{white}{Causation}}\\
ML learns correlation: decisions based on (possibly spurious) correlations and confounding (e.g., causing LLMs hallucinations) can be wrong and potentially dangerous.	
 & 
  The quality of the causal inference is superior, as it subsumes ML and offers a free-from-confounders decisions by ``cleaning'' the X-W relation from spurious correlations.
 \\ 
\
\cellcolor{blue!50}\textbf{\textcolor{white}{Reactive}} & \cellcolor{blue!50}\textbf{\textcolor{white}{Proactive}}\\
ML is a passive learning approach (i.e,. learning ``by seeing''), as it needs to wait for relevant knowledge to be observed. 
Strictly speaking, it learns the conditional probability of an event of interest $X$, given the observation of the event W, $P(X|W)$. &

Causal reasoning enables proactive inference by allowing hypotheses about outcomes under deliberate changes to variables. Through interventions, it supports planning and simulating \textit{improvement actions} beyond what is seen in training data. Formally, it learns quantities like  $P(X|do(W))$, where the do-operator indicates that 
$W$ is actively set rather than conditioned on other variables.
\\
\cellcolor{blue!50}\textbf{\textcolor{white}{Opaqueness}} & \cellcolor{blue!50}\textbf{\textcolor{white}{Transparency}}\\
The ML state-of-the-art strategy, that is deep learning, is opaque and lacks explainability, a severe limitations in safety-critical, legal or healthcare domains.  %
&
 Explainability and accountability is central to decision-making. Causal models provides fully explainable and transparent reasoning paths along with the result. %
\\
    \cellcolor{blue!50}\textbf{\textcolor{white}{Implicit}} & \cellcolor{blue!50}\textbf{\textcolor{white}{Explicit}}\\
ML-based solutions are purely data-driven. They do not integrate human knowledge and assumptions about the modelled phenomenon, which are essential to decision-making. Thus, context is assumed to be implicitly described by the data. This does not mean there are no assumptions, but that assumptions are not made explicit. &%

A causal model allows to easily specify the domain-dependent assumptions and integrate human knowledge in the loop. This improves the quality of decisions and gives a clearer framework, in terms of assumptions and constraints (i.e., context) under which those decisions are to be considered valid. 
\end{tabularx}
}
\label{RBSE:DDSE}
\vspace{-3pt}
\end{tcolorbox}

\section{The CSE conceptual framework}

We outline a conceptual framework to harness causal reasoning for software engineering tasks ranging from requirements analysis to design-space exploration, test generation, code analysis, and root cause analysis. In the next section, we review recent research efforts, including our own work on testing and analysis, that instantiate elements of this framework. 
 
\begin{wrapfigure}{r}{0.65\textwidth}
    \centering
    \vspace{-10pt}
    \includegraphics[width=\linewidth]{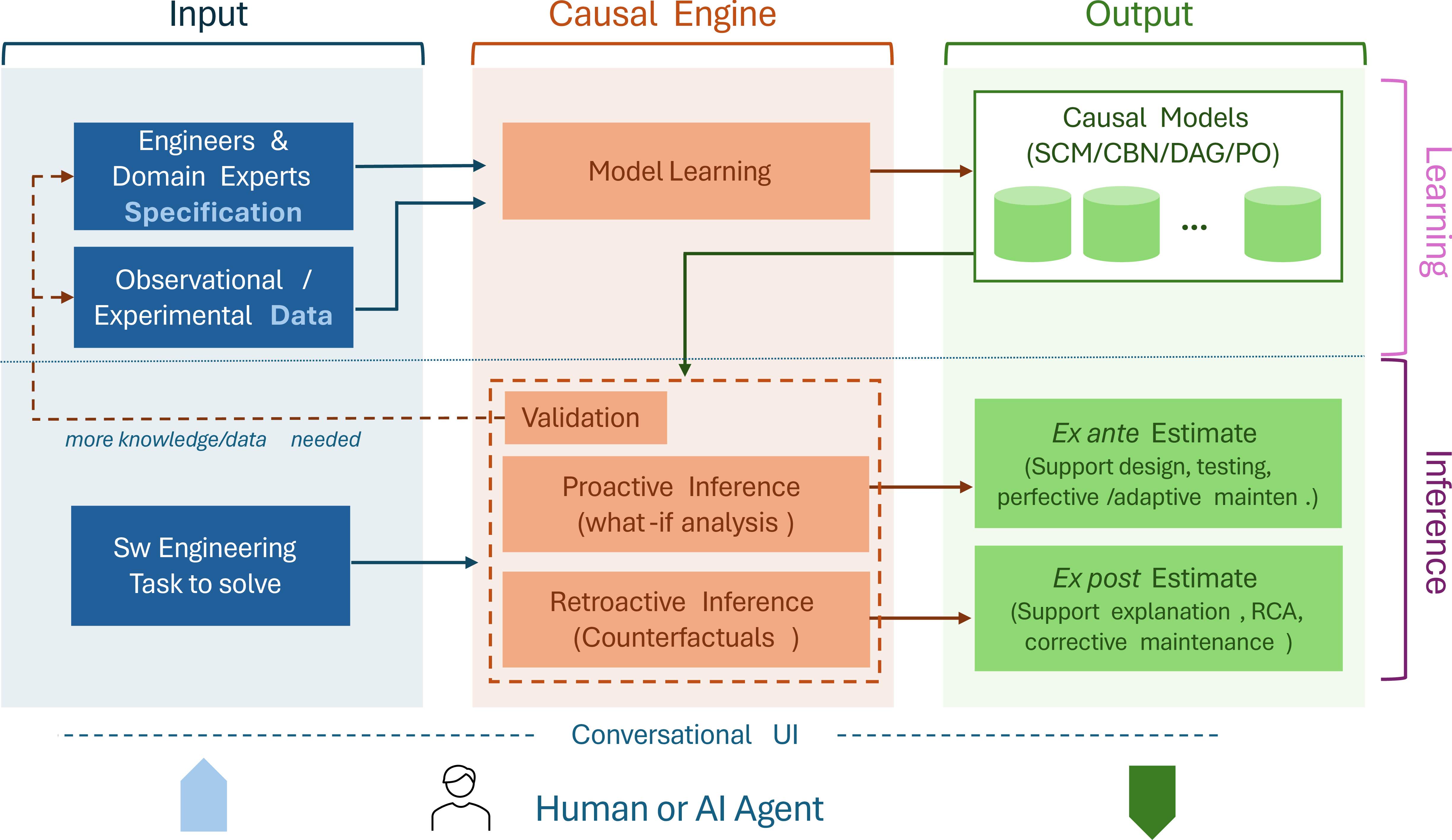}
    \vspace{-20pt}
    \caption{The CSE conceptual framework}
    \label{fig:RBSE}
    \vspace{-6pt}
\end{wrapfigure}

The framework (Figure \ref{fig:RBSE}) revolves around \textbf{causal models}: structured, inspectable representations of a system and its environment that encode the cause–effect relations relevant to a given task. Different tasks may call for different models; Table \ref{models} provides illustrative variables.
A causal model is a mathematical representation of causal relationships within a system. There are two main approaches to causal modeling and inference: the \textit{potential outcomes} framework, which frames causality in terms of what happened with what would have happened under an alternative condition; and \textit{graphical models}, where relationships are expressed through \textit{directed acyclic graphs}, with associated conditional probabilities (in causal Bayesian networks) or structural equations (in structural causal models) to specify how variables influence one another \cite{Imbens20, Pearl18}. We focus on graphical models in the description, though the core ideas generalize across formalisms.

Although there are many ways to instantiate a CSE solution for a specific problem, the idea is that \textit{causal models} should be at the core: they must be used to drive and check the reasoning process to solve the task \textit{rigorously} and \textit{transparently}. Rigorously means capable of quantifying the causal effects along the whole reasoning chain and harnessing such estimates to give the best output (e.g., generate safety tests more likely to cause a mishap). Transparently means giving inspectable evidence of the causal paths followed, along with the underpinning assumptions, constraints, and used sources of information, to justify the output for the required input, so that engineers are always kept in the loop and can challenge, refine, or override the reasoning when needed. In this paradigm, the causal model acts as a collaborative reasoning partner, not a black-box oracle: models help engineers reason more systematically, while engineers can encode those assumptions and constraints, such as required or forbidden causal links, that no automated method can reliably infer.
There are two steps to implement a CSE solution (Figure \ref{fig:RBSE}): 1) How to construct a causal model (\textbf{Learning}); 2) How to query the built model (\textbf{Inference}). %

\setlength{\arrayrulewidth}{0.5mm}
\setlength{\tabcolsep}{1pt}

\begin{table*}[t]
    \caption{Examples of cause and effect variables, adapted from existing works 
    \vspace{-6pt}
    \cite{Mascia2025,Giamattei2024,Giamattei2025,Chen2024}}
    \label{models}
    \centering
    \resizebox{\textwidth}{!}{ 
   \footnotesize{   
    \arrayrulecolor{white} 
    \begin{tabularx}{\textwidth}{p{1,82cm}|>{\raggedright\arraybackslash}p{2,8cm}|>{\raggedright\arraybackslash}X|>{\raggedright\arraybackslash}p{2,5cm}}
	\hline
	\cellcolor{orange!70}\textcolor{white}{\textbf{System}}  &  \cellcolor{orange!70}\textcolor{white}	{\textbf{Effect}} & 	\cellcolor{orange!70}\textcolor{white}{\textbf{Causes} }& \cellcolor{orange!70}\textcolor{white}{\textbf{Example Task}}
	 \\ \hline 
	\cellcolor{orange!10}\textcolor{brown!30!black}{Microservice Architecture}& \cellcolor{orange!10}\textcolor{brown!30!black}{Performance metrics (e.g., mean response time, throughput)} &\cellcolor{orange!10}\textcolor{brown!30!black}{Request rate and type per service, CPU and memory usage per service, latency between services, service topology (invocation matrix), load balancing strategy, instance replication}&\cellcolor{orange!10}\textcolor{brown!30!black}{ Evaluate performance impact of different deployment and scaling configurations}\\\hline 
	\cellcolor{orange!10}\textcolor{brown!30!black}{Autonomous \newline Driving System}&\cellcolor{orange!10}\textcolor{brown!30!black}{ Safety metrics (e.g., minimum distance to pedestrian, time-to-collision)}&\cellcolor{orange!10}\textcolor{brown!30!black}{Weather conditions (rain, fog, snow), road type (urban, rural, highway), ego vehicle speed, braking system response time, pedestrian behavior (speed, trajectory), sensor accuracy/degradation}&\cellcolor{orange!10}\textcolor{brown!30!black}{ Generate safety-critical test cases for varying driving scenarios}\\\hline 
	\cellcolor{orange!10}\textcolor{brown!30!black}{Loan Granting \newline System} &\cellcolor{orange!10}\textcolor{brown!30!black}{Fairness metrics (e.g., Statistical Parity Difference, Equal Opportunity Difference)} &\cellcolor{orange!10}\textcolor{brown!30!black}{Applicant socio-economic features (e.g., job type, income, education), financial history, credit score, protected attributes (e.g., sex, race, age), decision-making policy thresholds} &\cellcolor{orange!10}\textcolor{brown!30!black}{Assess and mitigate bias in loan approval decisions} 		
\end{tabularx}
}
}
\vspace{-6pt}
\end{table*}

\subsection{Causal Model Learning}
Causal models can be specified \textit{i)} manually by domain experts, \textit{ii)} automatically from data (either observational or from controlled experiments), or \textit{iii)} combining both approaches. Accordingly, the inputs to the model learning step system are a \textbf{specification} and/or \textbf{data}.

In CSE, a \textbf{specification} encodes the explicit domain knowledge relevant to the task. 
This knowledge can be used to fully define a model or to refine and complement an existing model generated from data. In both scenarios, experts express assumptions through cause–effect relationships, declaring which relations must hold and which are impossible, defining the context within which causal reasoning is valid. 
This manual step embeds human understanding directly into the model, allowing engineers to shape and constrain the reasoning space according to their expertise.
 
For instance, suppose we are designing or testing an emergency braking subsystem in an autonomous vehicle. 
A causal model is meant to help explore design alternatives and/or generate meaningful test cases that highlight a causal pathway from inputs to the possibly failing outputs. A domain expert may provide a causal model expressing key cause-effect relationships, such as: 
\begin{itemize}[leftmargin=*]
    \item Rain or fog (\texttt{[Weather]}) increases \texttt{[Sensor Noise]};
    \item Slippery road (\texttt{[Road Surface]}) increases \texttt{[Braking Distance]};
    \item High \texttt{[Vehicle Speed]} reduces \texttt{[Object Detection Accuracy]} due to motion blur;
    \item  Low \texttt{[Object Detection Accuracy]} may affect the \texttt{[Braking Decision]}. %
\end{itemize}

These expert-specified causal relations 
serve as an explicit, inspectable record of assumptions — something ML models do not provide.
Modern LLMs can complement human knowledge by suggesting additional cause–effect variables or relationships  
and build causal models \cite{wan2025,kiciman2024}. Further variables or relations could be elicited by an LLM, for instance: 

\begin{itemize}[leftmargin=*]
    \item \texttt{[Lighting Conditions]} $\rightarrow$ \texttt{[Sensor Noise]};
    \item \texttt{[Vehicle Load]} $\rightarrow$ \texttt{[Braking Distance]};
    \item \texttt{[Software Version]} $\rightarrow$ \texttt{[Braking Algorithm Latency]}
\end{itemize}
 These relations, parameterised by data (see below), are then exploited in the inference phase (Sec.~ \ref{causalinference}) to run queries and assess the causal effects of hypothetical interventions on variables. %

\textbf{Data} play a critical role in causal learning, in at least these ways:
\textit{(1) Causal structure discovery} methods can automatically extract cause–effect relationships from data, either entirely from observations or, more effectively, in combination with domain expertise expressed through a specification like above. Engineers can examine, validate, and revise these learned structures, closing the loop between automated discovery and human oversight.
A variety of mature approaches exist \cite{wang24_cd}, which are able to infer which variables influence others and in what direction. Importantly, modern solutions can learn causal structures not only from structured data (e.g., tabular monitoring metrics) but also from unstructured sources such as requirements, design documents, and natural-language logs,  leveraging again LLMs to extract causal relationships directly from text \cite{liu2024}.
Such capabilities make it possible to formalize  any type of knowledge about the system or process into an explicit causal model. 
\textit{(2) Fitting}. However the causal structure is obtained (from data or expert specification), data remain essential to \textit{quantify} the relationships — for example, to parameterize structural equations in an SCM or conditional distributions in a CBN. 
\textit{(3) Validation}. Data also enable validation of assumptions: if observations contradict a human-specified relation, the modeler can revisit whether the assumption is wrong or whether the data are simply insufficient \cite{Pearl18}.

\subsection{Causal Inference}
\label{causalinference}
With a model in hand, engineers can query it in a proactive mode — asking \textit{what-if} questions — or retroactively through counterfactual \textit{what-would-have-happened-if} questions, to simulate scenarios that may impact the effect variable(s) of interest. %
Continuing the self-driving car example, engineers can test \textit{What is the effect of a combined change in \texttt{[Weather]} and \texttt{[Road Surface]} on \texttt{[Braking Distance]}}; \textit{How degraded \texttt{[Object Detection Accuracy]} affects \texttt{[Braking Decision]} under different \texttt{[Vehicle Speeds]}}; or they can run counterfactual root cause analysis such as: \textit{Would a collision have been avoided if the \texttt{[Software Version]} had been different?} 
As a further illustration, consider a microservice system where engineers aim to improve the \texttt{[latency]} of the \textit{Checkout} service. A causal model, learned from monitoring data (logs, metrics, traces) and optionally refined by engineers, may relate latency to factors such as request rate, CPU and memory usage, inter-service latency (e.g., calls to \textit{Payment} or \textit{Inventory}), service topology (synchronous vs.\ asynchronous calls), load-balancing strategy, number of replicas. Typical dependencies might include: higher \texttt{[request rate]} $\rightarrow$ higher \texttt{[CPU]} $\rightarrow$ higher \texttt{[latency]}; more \texttt{[replicas]} $\rightarrow$ lower per-instance \texttt{[CPU]}; \texttt{[load-balancing strategy]} $\rightarrow$ \texttt{[CPU]} $\rightarrow$ \texttt{[latency]}.
With such a model, engineers can pose what-if questions: \textit{What if we increase \texttt{[replicas]}? What if we change \texttt{[load-balancing]}? What if a synchronous call becomes asynchronous?}, predicting causal effects on \texttt{[latency]}. Unlike ML dashboards, which can be distorted by confounding factors (e.g., ``more replicas cause higher latency'' simply because replicas increase under high load), a causal model supports true intervention analysis, helping engineers simulate actions, isolate confounders, and anticipate their impact.

There are multiple ways to instantiate this step, depending on the task and the chosen causal reasoning framework. The most direct approach is to estimate the causal effect of action — either those to take or those that could have been taken — to support tasks such as performance prediction, test generation, or root cause analysis.\footnote{Guo \textit{et al.} review and classify methods for estimating causal effects \cite{Guo2020}.}
With existing tools such as \texttt{PyWhy},\footnote{\url{https://www.pywhy.org/}}
 causal effect estimates can be produced with confidence intervals and validated through techniques like \textit{refutation tests}.

Alternatively, causal models can be combined with other reasoning components, such as LLM-powered AI agents capable of \textit{approximate reasoning} \cite{liu2025,kiciman2024}. Recent work shows how a causal model can guide an LLM in generating performance-critical configurations for microservice testing \cite{Mascia2025}. In such setups, the causal model may inject causal knowledge into the LLM’s context or act as a consistency checker 
to ensure outputs respect established causal relations.  %
In some settings, the aim is not to estimate causal effects but simply to uncover cause–effect relationships. For instance, in root cause analysis, causal models often help traverse the graph to identify likely failure causes, sometimes using heuristics such as weighted edges to approximate influence \cite{Wu21}.

The instantiation of a CSE solution depends on several factors: the available inputs (specification and structured/unstructured data), the SE task, the type of causal model (from simple graphs to fully parameterized models), the chosen inference approach (effect estimation, attribution, counterfactuals, graph traversal), and whether it is combined with tools such as LLMs. We next review how causal methods are currently used in software engineering.

\section{Where we are, what's next}

\subsection{The growing footprint of Causal Reasoning in SE}
Recent review studies including our own \cite{Giamattei2025,Siebert23} show a growing academic interest in applying causal methods to SE, with more than 90 papers mapped across major digital libraries. With few exceptions, most existing works can be viewed as instances of the CSE framework, typically applying a specific causal learning or inference method to an SE problem. %
These studies highlight the benefits of combining reasoning with data, especially when both causal discovery and inference are used.
\begin{figure*}[t!]
    \centering
    \includegraphics[width=0.8\textwidth]{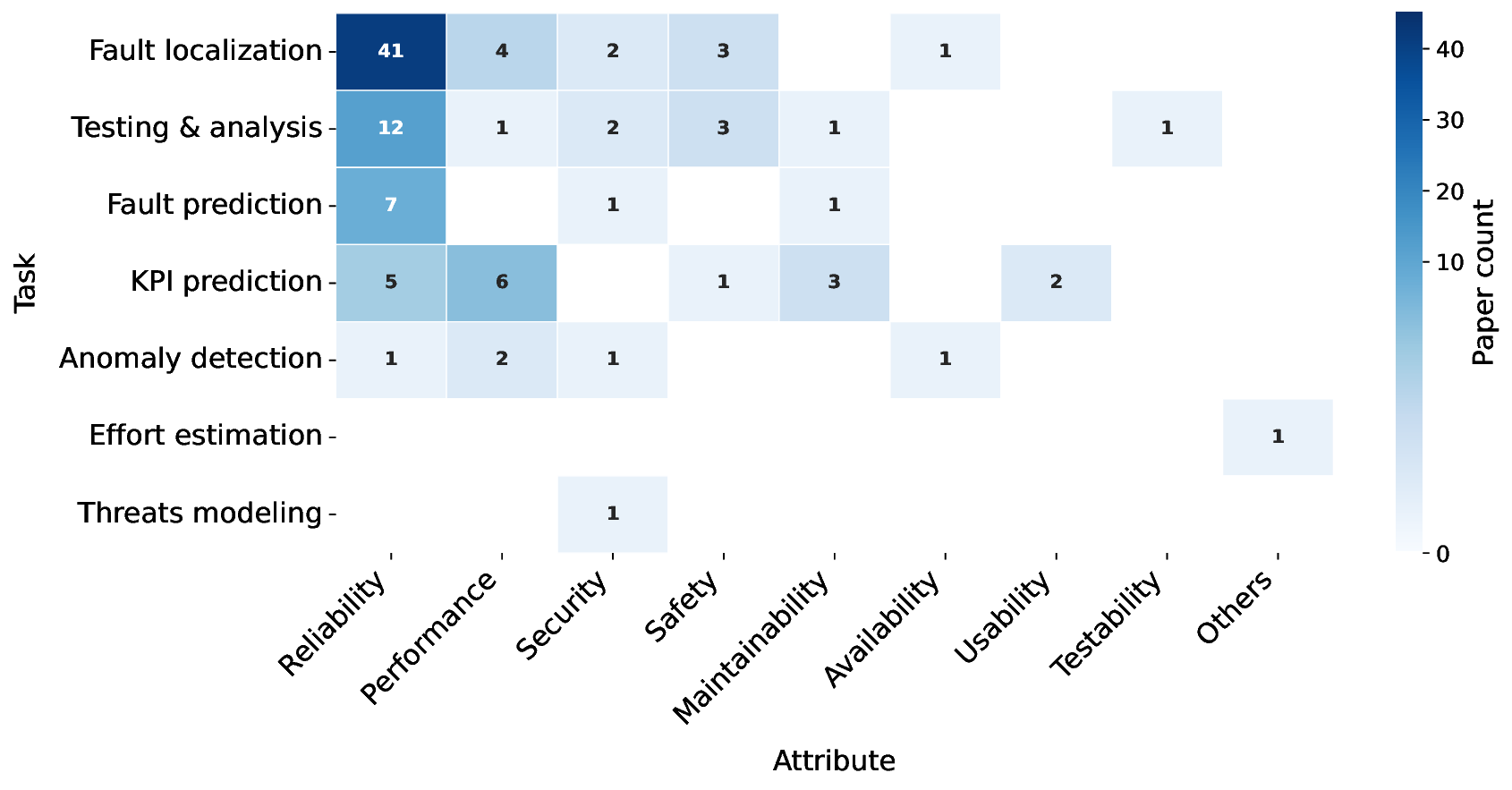}
    \vspace{-6pt}
    \caption{Causal methods in software engineering from the intersection of mapping studies \cite{Giamattei2025, Siebert23}}
    \label{fig:sota}
    \vspace{-6pt}
\end{figure*}
\textbf{Main areas of application} (Figure \ref{fig:sota}) %
include: 
\textit{(1) Fault localization}: 
Causal graphs are used to model distributed systems and trace failure-propagation chains, especially in microservices. Most work focuses on learning causal structures (e.g., service-dependency graphs), while true causal inference -- such as counterfactual root cause analysis — remains uncommon. Extracted graphs are often handed to ranking or graph-traversal algorithms, e.g., PageRank \cite{Wu21}. In debugging, however, causal inference is more prominent, with models derived from programs used for statistical fault localization and program repair \cite{Johnson2020}.
\textit{(2) Testing} is naturally causal: choosing an input implicitly involves predicting outcomes under hypothetical conditions. CSE operationalizes this reasoning. It has been applied to self-driving cars \cite{Giamattei2024}, microservices \cite{Giamattei2024_2,Mascia2025}, cyber-physical systems \cite{foster2025}, and metamorphic testing  \cite{Clark2023}. 
\textit{(3) Prediction:} 
Causal modeling enables predictions free from confounding, unlike correlational ML. Applications include estimating design factors (e.g., reuse success \cite{Gang2024}), maintenance (refactoring) impact \cite{Hamdi2021}, fault prediction \cite{Hu2023}. %
Although relatively few works implement a full discovery and inference pipeline, the range of causal applications is expanding. Opportunities also exist earlier in the lifecycle, such as requirements elicitation and design exploration, where what-if reasoning is highly valuable.

\noindent \textbf{Quality attributes.} 
Most work targets reliability, some performance. Applications to security and safety remain rare, despite causal models' explanatory power is potentially useful for certification.

\noindent\textbf{Tools.} 
Causal discovery and inference tooling is maturing rapidly. The \texttt{PyWhy} ecosystem (e.g., \texttt{causal-learn}, \texttt{dowhy}, \texttt{EconML}), IBM’s \texttt{SPSS Amos}\footnote{\url{https://www.ibm.com/it-en/products/structural-equation-modeling-sem}}
, and the \texttt{Tetrad} toolbox\footnote{\url{https:///github.com/cmu-phil/tetrad}} are notable examples. While these are general-purpose tools, the growth of CSE is likely to spur SE-specific extensions (e.g., for testing or debugging), and conversely, SE practices may help improve causal tooling, echoing the two-way synergy already observed between ML and SE. 

\subsection{Practical Challenges and Pathways to Adoption}
Despite the promise shown in the literature, CSE is not yet ready to supplant today’s DDSE solutions. Its adoption in industry will depend on addressing several practical challenges

\noindent \textbf{Expertise barriers}. Although causal models are interpretable, using them effectively still requires familiarity with causal thinking and with some statistical foundations (e.g., structure discovery, model fitting, causal inference). Most engineering teams today have limited exposure to these concepts.
\textit{Pathway to adoption}. 
Promote incremental adoption through tools that scaffold causal thinking, such as  \texttt{PyWhy}, supported by tutorials, examples, and domain-specific case studies.

\noindent \textbf{Data quality and models stability}. 
 As with ML, causal inference relies on data that sufficiently capture the relevant variables. Yet many SE datasets — logs, metrics, traces, issues, commits — are noisy, incomplete, sparse, or biased. Unlike physical systems, software behavior is driven by discrete events, rare failures, and non-stationary processes. Consequently, causal structures inferred purely from observational SE data can be unstable, limiting their generalizability.
\textit{Pathway to adoption}. 
Establish robustness checks and refutation tests as standard practice, a procedures still largely missing in today’s SE causal literature \cite{Menzies2025}. These include falsification tests, placebo tests (detecting effects where none should exist), subset refutation (testing consistency across time windows), simulated confounders (sensitivity to missing variables), and bootstrapping for stability assessment. Such practices, supported by libraries like \texttt{DoWhy}, are essential for determining whether causal conclusions are trustworthy enough for engineering use.

\noindent \textbf{Integration with existing workflows and tools}. 
Most organizations depend on established CI/CD pipelines, observability stacks, testing frameworks, and debugging tools. Introducing causal models requires integration rather than replacement.
\textit{Pathway to adoption}. 
Develop CSE plug-ins for existing SE tools (e.g., test generators, performance profilers, failure analyzers) so causal reasoning enhances current practices. Like ML, causal reasoning consumes data, but it allows engineers to guide, constrain, and interpret outputs in causal terms. Positioning CSE as an augmentation layer atop mature ML tooling will significantly increase the likelihood of industrial adoption.

\noindent \textbf{Computational and tooling maturity}. 
Although causal inference libraries are improving, they are not yet as turnkey, scalable, or optimized as mainstream ML toolkits. Inference can be computationally heavy, and automated discovery requires careful validation.
\textit{Pathway to adoption}. 
Invest in efficient, domain-specific, SE-tailored implementations, cloud-based causal services, and high-level abstractions that hide complexity behind intuitive interfaces.

\noindent \textbf{Organizational resistance and cultural change}. 
Adopting causal modeling requires a shift in how engineers reason about systems. Many teams, accustomed to correlational ML tools, may view causal modeling as too theoretical or demanding.
\textit{Pathway to adoption}. 
Highlight success stories and industrial case studies that show tangible benefits, e.g., in terms of fewer failures, more effective tests, clearer explanations. Demonstrated value will be key to cultural adoption.

\subsection{Long-term evolution}
\noindent \textbf{Reasoning- and data-driven integration.} 
Beyond its immediate utility in SE tasks, CSE's broader promise lies in integrating causal reasoning with data. This shift is already visible in research on LLMs’ emerging reasoning abilities \cite{Plaat25} and  
in LLM-powered AI agents \cite{liu2025}, enabling planning and improving explainability.  %
As AI (multi)agent systems such as MetaGPT\footnote{\url{https://github.com/FoundationAgents/MetaGPT}}
 gain traction in software engineering, 
 CSE is well positioned to support these emerging architectures.

\noindent \textbf{Logic besides statistics.}
While we assumed \textit{statistical} causal inference, there are many flavours of causality. %
For instance, in law and forensic analysis, logic-based causality dominates, relying on formal rules and domain knowledge \cite{kiciman2024}. %
 Linguistic causality — based on semantic relations in text — is also valuable, especially when working with natural-language artifacts such as requirements. A comprehensive CSE framework should treat these different forms of causality as first-class citizens.

\noindent \textbf{Human in-the-loop.}  
One key strength of causal models is their interpretability, making CSE naturally suited for human-machine co-design. Inspectable models support grounded, trustworthy responses, and experts can refine them by adding or removing causal links.  In CSE, engineers are collaborators rather than spectators; we depart from any ``push-button'' vision of fully autonomous SE, as trustworthy systems cannot be built without humans in the loop.

\noindent \textbf{Ethical concerns} around AI must not be overlooked. Beyond well-known issues (e.g., safety, privacy, fairness) now under regulatory scrutiny, a subtler risk is the erosion of human judgment and creativity as AI systems grow more capable. In SE, powerful assistants may marginalize human decision-making. An emerging counterpoint is the view of \textit{AI as a provocateur} rather than a servant \cite{Advait2024}: a partner that challenges assumptions, surfaces biases, and stimulates critical thinking. The transparency of causal models naturally supports this reflective engagement. %
We fully embrace this view:  AI can augment human intelligence only when humans remain actively involved.%
\newpage

\bibliographystyle{ACM-Reference-Format}
\bibliography{acmart}

@article{Hou2024,
author = {Hou, Xinyi and Zhao, Yanjie and Liu, Yue and Yang, Zhou and Wang, Kailong and Li, Li and Luo, Xiapu and Lo, David and Grundy, John and Wang, Haoyu},
title = {Large Language Models for Software Engineering: A Systematic Literature Review},
year = {2024},
issue_date = {November 2024},
publisher = {Association for Computing Machinery},
address = {New York, NY, USA},
volume = {33},
number = {8},
issn = {1049-331X},
url = {https://doi.org/10.1145/3695988},
doi = {10.1145/3695988},
journal = {ACM Trans. Softw. Eng. Methodol.},
month = dec,
articleno = {220},
numpages = {79}
}

@inproceedings{Battaglini2025, series={ICPE ’25},
   title={FAILS: A Framework for Automated Collection and Analysis of LLM Service Incidents},
   url={http://dx.doi.org/10.1145/3680256.3721320},
   DOI={10.1145/3680256.3721320},
   booktitle={Companion of the 16th ACM/SPEC International Conference on Performance Engineering},
   publisher={ACM},
   author={Battaglini-Fischer, Sándor and Srinivasan, Nishanthi and Szarvas, Bálint László and Chu, Xiaoyu and Iosup, Alexandru},
   year={2025},
   month=may, pages={187–194},
   collection={ICPE ’25} }

@article{tie2024_tosem,
author = {Tie, Jiessie and Yao, Bingsheng and Li, Tianshi and Fang, Hongbo and Ahmed, Syed Ishtiaque and Wang, Dakuo and Zhou, Shurui},
title = {"Should I Give Up Now?" Investigating LLM Pitfalls in Software Engineering},
year = {2026},
publisher = {Association for Computing Machinery},
address = {New York, NY, USA},
issn = {1049-331X},
doi = {10.1145/3801972},
journal = {ACM Trans. Softw. Eng. Methodol.},
month = mar
}

@ARTICLE{Wang2023,
  author={Wang, Simin and Huang, Liguo and Gao, Amiao and Ge, Jidong and Zhang, Tengfei and Feng, Haitao and Satyarth, Ishna and Li, Ming and Zhang, He and Ng, Vincent},
  journal={IEEE Transactions on Software Engineering}, 
  title={Machine/Deep Learning for Software Engineering: A Systematic Literature Review}, 
  year={2023},
  volume={49},
  number={3},
  pages={1188-1231},
  doi={10.1109/TSE.2022.3173346}}

@INPROCEEDINGS {huang2025,
author = { Huang, Kaicheng and Wang, Fanyu and Huang, Yutan and Arora, Chetan },
booktitle = {33rd International Requirements Engineering Conference Workshops},
title = {{ Prompt Engineering for Requirements Engineering: A Literature Review and Roadmap }},
year = {2025},
publisher = {IEEE},
volume = {},
ISSN = {},
pages = {548-557},
doi = {10.1109/REW66121.2025.00081},
publisher = {IEEE Computer Society},
address = {Los Alamitos, CA, USA}
}

@article{jiang2024,
author = {Jiang, Juyong and Wang, Fan and Shen, Jiasi and Kim, Sungju and Kim, Sunghun},
title = {A Survey on Large Language Models for Code Generation},
year = {2025},
publisher = {ACM},
address = {New York, NY, USA},
issn = {1049-331X},
doi = {10.1145/3747588},
journal = {ACM Trans. Softw. Eng. Methodol.}
}

@ARTICLE{Wang2024,
  author={Wang, Junjie and Huang, Yuchao and Chen, Chunyang and Liu, Zhe and Wang, Song and Wang, Qing},
  journal={IEEE Transactions on Software Engineering}, 
  title={Software Testing With Large Language Models: Survey, Landscape, and Vision}, 
  year={2024},
  volume={50},
  number={4},
  pages={911-936},
  doi={10.1109/TSE.2024.3368208}}

@article{zhang2024_tosem,
author = {Zhang, Quanjun and Fang, Chunrong and Xie, Yang and Ma, Yuxiang and Sun, Weisong and Yang, Yun and Chen, Zhenyu},
title = {A Systematic Literature Review on Large Language Models for Automated Program Repair},
year = {2026},
publisher = {ACM},
address = {New York, NY, USA},
issn = {1049-331X},
doi = {10.1145/3799693},
journal = {ACM Trans. Softw. Eng. Methodol.},
month = mar
}

@inproceedings{Majdoub2024,
author = {Majdoub, Yacine and Ben Charrada, Eya},
title = {Debugging with Open-Source Large Language Models: An Evaluation},
year = {2024},
isbn = {9798400710476},
publisher = {Association for Computing Machinery},
address = {New York, NY, USA},
url = {https://doi.org/10.1145/3674805.3690758},
doi = {10.1145/3674805.3690758},
booktitle = {Proceedings of the 18th ACM/IEEE International Symposium on Empirical Software Engineering and Measurement},
pages = {510–516},
numpages = {7},
location = {Barcelona, Spain},
series = {ESEM '24}
}

@book{Pearl18,
author = {Pearl, Judea and Mackenzie, Dana},
title = {The Book of Why: The New Science of Cause and Effect},
year = {2018},
isbn = {046509760X},
publisher = {Basic Books, Inc.},
address = {USA},
edition = {1st}
}

@article{kiciman2024,
title={Causal Reasoning and Large Language Models: Opening a New Frontier for Causality},
author={Emre Kiciman and Robert Ness and Amit Sharma and Chenhao Tan},
journal={Transactions on Machine Learning Research},
pages={2835-8856},
year={2025},
url={https://openreview.net/forum?id=mqoxLkX210}
}

@inproceedings{wan2025,
author = {Wan, Guangya and Lu, Yunsheng and Wu, Yuqi and Hu, Mengxuan and Li, Sheng},
title = {Large language models for causal discovery: current landscape and future directions},
year = {2025},
isbn = {978-1-956792-06-5},
doi = {10.24963/ijcai.2025/1186},
booktitle = {34th International Joint Conference on Artificial Intelligence},
articleno = {1186},
numpages = {9},
location = {Montreal, Canada},
series = {IJCAI '25}
}

@inproceedings{liu2025,
    title = "Large Language Models and Causal Inference in Collaboration: A Comprehensive Survey",
    author = "Liu, Xiaoyu  and Xu, Paiheng  and Wu, Junda  and Yuan, Jiaxin  and Yang, Yifan  and Zhou, Yuhang  and Liu, Fuxiao  and Guan, Tianrui  and Wang, Haoliang  and Yu, Tong  and McAuley, Julian  and Ai, Wei  and Huang, Furong",
    editor = "Chiruzzo, Luis  and Ritter, Alan  and Wang, Lu",
    booktitle = "Findings of the Association for Computational Linguistics: NAACL 2025",
    month = apr,
    year = "2025",
    address = "Albuquerque, New Mexico",
    publisher = "ACL",
    doi = "10.18653/v1/2025.findings-naacl.427",
    pages = "7668--7684",
    ISBN = "979-8-89176-195-7"
}

@article{Imbens20,
	author = {Imbens, Guido W.},
	doi = {10.1257/jel.20191597},
	journal = {Journal of Economic Literature},
	number = {4},
	pages = {1129-79},
	title = {{Potential Outcome and Directed Acyclic Graph Approaches to Causality: Relevance for Empirical Practice in Economics}},
	volume = {58},
	year = {2020}}

@inproceedings{liu2024,
title={Discovery of the Hidden World with Large Language Models},
author={Chenxi Liu and Yongqiang Chen and Tongliang Liu and Mingming Gong and James Cheng and Bo Han and Kun Zhang},
booktitle={38th Annual Conference on Neural Information Processing Systems},
year={2024},
url={https://openreview.net/forum?id=w50ICQC6QJ}
}

@article{Guo2020,
	articleno = {75},
	author = {Guo, Ruocheng and Cheng, Lu and Li, Jundong and Hahn, P. Richard and Liu, Huan},
	doi = {10.1145/3397269},
	issn = {0360-0300},
	issue_date = {July 2021},
	journal = {ACM Computing Surveys},
	number = {4},
	numpages = {37},
	publisher = {ACM},
	title = {{A Survey of Learning Causality with Data: Problems and Methods}},
	volume = {53},
	year = {2020}}

@INPROCEEDINGS{Mascia2025,
  author={Mascia, Cristian and Guerriero, Antonio and Giamattei, Luca and Pietrantuono, Roberto and Russo, Stefano},
  booktitle={2025 IEEE/ACM Second International Conference on AI Foundation Models and Software Engineering (Forge)}, 
  title={Microservices Performance Testing with Causality-enhanced Large Language Models}, 
  year={2025},
  volume={},
  number={},
  pages={136-140},
  doi={10.1109/Forge66646.2025.00022}}

@article{Giamattei2025,
	author = {Luca Giamattei and Antonio Guerriero and Roberto Pietrantuono and Stefano Russo},
	doi = {https://doi.org/10.1016/j.infsof.2024.107599},
	issn = {0950-5849},
	journal = {Information and Software Technology},
	pages = {107599},
	title = {Causal reasoning in Software Quality Assurance: A systematic review},
	url = {https://www.sciencedirect.com/science/article/pii/S0950584924002040},
	volume = {178},
	year = {2025}}

@article{Siebert23,
	address = {USA},
	author = {Siebert, Julien},
	doi = {10.1016/j.infsof.2023.107198},
	issn = {0950-5849},
	issue_date = {Jul 2023},
	journal = {Information and Software Technology},
	number = {C},
	numpages = {16},
	publisher = {Butterworth-Heinemann},
	title = {Applications of Statistical Causal Inference in Software Engineering},
	volume = {159},
	year = {2023}}

@inproceedings{Wu21,
	author = {Wu, Li and Tordsson, Johan and Elmroth, Erik and Kao, Odej},
	booktitle = {IEEE International Conference on Autonomic Computing and Self-Organizing Systems (ACSOS)},
	doi = {10.1109/ACSOS52086.2021.00029},
	pages = {21-30},
	publisher = {IEEE},
	title = {Causal Inference Techniques for Microservice Performance Diagnosis: Evaluation and Guiding Recommendations},
	year = {2021}}

@inproceedings{Johnson2020,
	author = {Johnson, Brittany and Brun, Yuriy and Meliou, Alexandra},
	booktitle = {2020 ACM/IEEE 42nd International Conference on Software Engineering (ICSE)},
	doi = {10.1145/3377811.3380377},
	isbn = {9781450371216},
	location = {Seoul, South Korea},
	numpages = {13},
	pages = {87--99},
	publisher = {ACM},
	title = {Causal testing: understanding defects' root causes},
	year = {2020}}

@article{Giamattei2024,
	articleno = {74},
	author = {Giamattei, Luca and Guerriero, Antonio and Pietrantuono, Roberto and Russo, Stefano},
	doi = {10.1145/3635709},
	issn = {1049-331X},
	issue_date = {March 2024},
	journal = {ACM Transactions on Software Engineering and Methodology},
	number = {3},
	publisher = {ACM},
	title = {Causality-driven Testing of Autonomous Driving Systems},
	volume = {33},
	year = {2024}}

@INPROCEEDINGS{Giamattei2024_2,
  author={Giamattei, Luca and Guerriero, Antonio and Malavolta, Ivano and Mascia, Cristian and Pietrantuono, Roberto and Russo, Stefano},
  booktitle={2024 IEEE/ACM International Conference on Automation of Software Test (AST)}, 
  title={Identifying Performance Issues in Microservice Architectures through Causal Reasoning}, 
  year={2024},
  volume={},
  number={},
  pages={149-153},
  doi={}}

@inproceedings{Clark2023,
	author = {Clark, Andrew G. and Foster, Michael and Walkinshaw, Neil and Hierons, Robert M.},
	booktitle = {2023 IEEE Conference on Software Testing, Verification and Validation (ICST)},
	doi = {10.1109/ICST57152.2023.00023},
	pages = {153-164},
	title = {Metamorphic Testing with Causal Graphs},
	year = {2023}}

@article{Chen2024,
author = {Chen, Zhenpeng and Zhang, Jie M. and Hort, Max and Harman, Mark and Sarro, Federica},
title = {Fairness Testing: A Comprehensive Survey and Analysis of Trends},
year = {2024},
issue_date = {June 2024},
publisher = {Association for Computing Machinery},
address = {New York, NY, USA},
volume = {33},
number = {5},
issn = {1049-331X},
url = {https://doi.org/10.1145/3652155},
doi = {10.1145/3652155},
journal = {ACM Trans. Softw. Eng. Methodol.},
month = jun,
articleno = {137},
numpages = {59}
}

@article{Hu2023,
	author = {Yamin Hu and Wenjian Luo and Zongyao Hu},
	doi = {10.1016/j.engappai.2023.106187},
	issn = {0952-1976},
	journal = {Engineering Applications of Artificial Intelligence},
	pages = {106187},
	title = {A practical approach to explaining defect proneness of code commits by causal discovery},
	volume = {123},
	year = {2023}}

@article{Gang2024,
	author = {Li, Gang and Dai, Honghua},
	doi = {10.1142/S021819400400166X},
	journal = {International Journal of Software Engineering and Knowledge Engineering},
	number = {03},
	pages = {351-364},
	title = {What will affect software reuse: A causal model analysis},
	volume = {14},
	year = {2004}}

@inproceedings{Hamdi2021,
	author = {Hamdi, Oumayma and Ouni, Ali and AlOmar, Eman Abdullah and {\'O} Cinn{\'e}ide, Mel and Mkaouer, Mohamed Wiem},
	booktitle = {2021 IEEE/ACM 8th International Conference on Mobile Software Engineering and Systems (MobileSoft)},
	doi = {10.1109/MobileSoft52590.2021.00010},
	pages = {28-39},
	title = {An Empirical Study on the Impact of Refactoring on Quality Metrics in {Android} Applications},
	year = {2021}}

@article{Advait2024,
author = {Sarkar, Advait},
title = {AI Should Challenge, Not Obey},
year = {2024},
issue_date = {October 2024},
publisher = {Association for Computing Machinery},
address = {New York, NY, USA},
volume = {67},
number = {10},
issn = {0001-0782},
url = {https://doi.org/10.1145/3649404},
doi = {10.1145/3649404},
journal = {Commun. ACM},
month = sep,
pages = {18–21},
numpages = {4}
}

@inproceedings{foster2025,
    title = {Using causal inference to test systems with hidden and interacting variables: an evaluative case study},
    doi = {10.1145/3756681.3756967},
    author = {Michael Foster and Robert M. Hierons and Donghwan Shin and Neil Walkinshaw and Christopher Wild},
    publisher = {ACM},
    year = {2025},
    booktitle = {29th International Conference on Evaluation and Assessment in Software Engineering (EASE 2025)},
}

@ARTICLE{baltes2025_tse,
  author={Baltes, Sebastian and Speith, Timo and Chiteri, Brenda and Mohsenimofidi, Seyedmoein and Chakraborty, Shalini and Buschek, Daniel},
  journal={IEEE Transactions on Software Engineering}, 
  title={On the Need to Rethink Trust in AI Assistants for Software Development: A Critical Review}, 
  year={2026},
  volume={},
  number={},
  pages={1-18},
  doi={10.1109/TSE.2026.3659804}}

@article{Khati2025,
author = {Khati, Dipin and Liu, Yijin and Palacio, David N. and Zhang, Yixuan and Poshyvanyk, Denys},
title = {Mapping the Trust Terrain: LLMs in Software Engineering - Insights and Perspectives},
year = {2025},
publisher = {ACM},
address = {New York, NY, USA},
issn = {1049-331X},
doi = {10.1145/3771282},
journal = {ACM Trans. Softw. Eng. Methodol.}
}

@techreport{CNCF2024,
  author      = {{Cloud Native Computing Foundation (CNCF)} and {Linux Foundation Research}},
  title       = {CNCF Annual Survey 2024},
  institution = {Linux Foundation Research},
  year        = {2025},
  url         = {https://www.cncf.io/wp-content/uploads/2025/04/cncf_annual_survey24_031225a.pdf},
  note        = {Accessed: 2026-03-16. Report presenting results of the 2024 CNCF Annual Survey.}
}

@techreport{CNCF2023,
  author      = {{Cloud Native Computing Foundation (CNCF)} and {Linux Foundation Research}},
  title       = {CNCF Annual Survey 2023},
  institution = {Linux Foundation Research},
  year        = {2023},
  url         = {https://www.cncf.io/reports/cncf-annual-survey-2023/},
  note        = {Accessed: 2026-03-16.}
}

@misc{McKinsey2024,
  author       = {Singla, Alex and Sukharevsky, Alexander and Yee, Lareina and Chui, Michael},
  title        = {The State of {AI} in Early 2024: Gen {AI} Adoption Spikes and Starts to Generate Value},
  howpublished = {McKinsey \& Company},
  year         = {2024},
  url          = {https://www.mckinsey.com/capabilities/quantumblack/our-insights/the-state-of-ai-2024},
  note         = {Accessed: 2026-03-16.}
}

@article{Martinez22,
author = {Mart\'{\i}nez-Fern\'{a}ndez, Silverio and Bogner, Justus and Franch, Xavier and Oriol, Marc and Siebert, Julien and Trendowicz, Adam and Vollmer, Anna Maria and Wagner, Stefan},
title = {Software Engineering for AI-Based Systems: A Survey},
year = {2022},
issue_date = {April 2022},
publisher = {ACM},
address = {New York, NY, USA},
volume = {31},
number = {2},
issn = {1049-331X},
doi = {10.1145/3487043},
journal = {ACM Trans. Softw. Eng. Methodol.},
month = apr,
articleno = {37e},
numpages = {59}
}

@article{wang24_cd,
title = {A survey of causal discovery based on functional causal model},
journal = {Engineering Applications of Artificial Intelligence},
volume = {133},
pages = {108258},
year = {2024},
issn = {0952-1976},
doi = {10.1016/j.engappai.2024.108258},
author = {Lei Wang and Shanshan Huang and Shu Wang and Jun Liao and Tingpeng Li and Li Liu}
}

@article{Plaat25,
author = {Plaat, Aske and Wong, Annie and Verberne, Suzan and Broekens, Joost and Van Stein, Niki and B\"{a}ck, Thomas},
title = {Multi-Step Reasoning with Large Language Models, a Survey},
year = {2025},
issue_date = {April 2026},
publisher = {ACM},
address = {New York, NY, USA},
volume = {58},
number = {6},
issn = {0360-0300},
doi = {10.1145/3774896},
journal = {ACM Comput. Surv.},
month = dec,
articleno = {160},
numpages = {35}
}

@article{Menzies2025,
	author = {Hulse, Jeremy and Eisty, Nasir U. and Menzies, Tim},
	date = {2025/07/21},
	doi = {10.1007/s10664-025-10690-6},
	id = {Hulse2025},
	isbn = {1573-7616},
	journal = {Empirical Software Engineering},
	number = {5},
	pages = {142},
	title = {Shaky structures: The wobbly world of causal graphs in software analytics},
	url = {https://doi.org/10.1007/s10664-025-10690-6},
	volume = {30},
	year = {2025}}

@InProceedings{TemporalMicroservicesTD,
author="Paudel, Bhuwan
and Gonzalez-Huerta, Javier
and Zabardast, Ehsan",
editor="Scanniello, Giuseppe
and Lenarduzzi, Valentina
and Romano, Simone
and Vegas, Sira
and Francese, Rita",
title="Temporal Evolution of Architectural Complexity and Technical Debt in Microservices: An Exploratory Case Study",
booktitle="Product-Focused Software Process Improvement",
year="2026",
publisher="Springer Nature Switzerland",
address="Cham",
pages="285--302",
isbn="978-3-032-12089-2"
}

\end{document}